\documentclass[11pt]{article}
\usepackage{geometry}
\usepackage{amsmath,amsfonts,amssymb,graphicx,url}
\usepackage{hyperref}
\usepackage{amsmath,amsfonts}
\usepackage{algorithmic}
\usepackage{algorithm}
\usepackage{array}
\usepackage[caption=false,font=normalsize,labelfont=sf,textfont=sf]{subfig}
\usepackage{textcomp}
\usepackage{stfloats}
\usepackage{url}
\usepackage{verbatim}
\usepackage{graphicx}
\usepackage{cite}

\usepackage{multicol}
\usepackage{amssymb}

\newtheorem{definition}{Definition}%

\begin{document}

\title{\textbf{\Large{Stable Acoustic Relay Assignment with High Throughput via Lase Chaos-based Reinforcement Learning}}}

\author{Zengjing Chen, Lu Wang*, Chengzhi Xing
\thanks{
\textbf{Note:} This work has been accepted to appear in \textit{IEEE Wireless Communications Letters}, 2025. \\
\textcopyright~2025 IEEE. Personal use of this material is permitted. Permission from IEEE must be obtained for all other uses, in any current or future
media, including reprinting/republishing this material for advertising or promotional purposes, creating new collective works, for resale or redistribution to servers or lists, or reuse of any copyrighted component of this work in other works.
}}


\maketitle

\begin{abstract}
This study addresses the problem of stable acoustic relay assignment in an underwater acoustic network. Unlike the objectives of most existing literature, two distinct objectives, namely classical stable arrangement and ambiguous stable arrangement, are considered. To achieve these stable arrangements, a laser chaos-based multi-processing learning (LC-ML) method is introduced to efficiently obtain high throughput and rapidly attain stability. In order to sufficiently explore the relay's decision-making, this method uses random numbers generated by laser chaos to learn the assignment of relays to multiple source nodes. This study finds that the laser chaos-based random number and multi-processing in the exchange process have a positive effect on higher throughput and strong adaptability with environmental changing over time. Meanwhile, ambiguous cognitions result in the stable configuration with less volatility compared to accurate ones. This provides a practical and useful method and can be the basis for relay selection in complex underwater environments.
\end{abstract}


\section{Introduction}
Underwater Acoustic Networks (UANs) have gained significant attention from both industry and academia due to their indisputable advantages in improving link reliability, increasing system capacity, expanding transmission range and so on. Acoustic communication is most widely used underwater communication as sound wave is not absorbed by water so easily like electromagnetic wave and optical wave \cite{HY2006}. UANs typically consist of acoustic-linked seabed sensors, autonomous underwater vehicles, and ground stations that provide links to onshore control centers. Due to the battery-powered network nodes, shallow water acoustic channel characteristics, such as low available bandwidth and highly varying multi-path, maximizing throughput while minimizing consumption has become a very challenging task \cite{SSP2000}. Recent studies have discussed the challenges and opportunities of underwater cognitive communication \cite{L2014}, proposed cooperative automatic repeat request protocols for higher channel quality \cite{J2019}, and analyzed the impact of low transmission rates and long preambles on medium access control protocols \cite{ZP2015}.

Artificial intelligence (AI) has experienced significant growth in popularity in recent years, and many industries and research fields have explored its potential applications, including information theory, game theory, biological systems, and so on \cite{ML2014,NVH2012,SDM1997,XSL2020}. One important subfield of AI is reinforcement learning (RL), which has proven to be an effective tool for solving sequential decision-making problems, building intelligent systems, and so on. RL is a learning process that involves an agent interacting with an environment in order to achieve a goal. The agent explores the unknown environment and learns the optimal behavior to obtain maximum reward through trial and error. Unlike traditional learning methods, RL does not rely on a supervisor or a teacher to provide explicit instructions. Instead, the agent learns from the feedback it receives from the environment and adapts its behavior accordingly. This process is similar to natural learning processes where learning evolves through interactions and observation. Studies such as \cite{CW2018,MLL2021,KBC2018,NA2019} have shown the promising results that can be achieved using RL.

Reinforcement learning has found widespread application in decision-making and optimization within wireless sensor networks and UANs \cite{WLQ2020,LZD2019,CLH2013,VP2019,WLQ2019,SGZ2021}. In particular, RL has been applied to improve communication stability through underwater acoustic channels, a highly variable and unpredictable environment. The core of the problem lies a trade-off between exploring poorly characterized locations and exploiting known ones, as described in \cite{CLH2013}, which uses a modified Gittins index rule to account for switching costs and includes field experiments in the Charles River Boston. Li at el. \cite{LLYH2016} achieved the relay selection which enables highly stable performance of the cooperative system in a complicated underwater environment. Additionally, \cite{LLYH2018} introduces a multi-user multi-armed bandit (MU-MAB) approach to address the challenge faced by UANs, where users lack prior knowledge of underwater acoustic channel conditions. In \cite{ZZC2021}, a RL-based opportunistic routing protocol is proposed by combining the advantages of opportunistic routing and RL algorithm. Furthermore, Geng at el. \cite{GZ2022} proposes a novel deep-RL-based medium access control protocol (DL-MAC) for UANs where one agent node employing the proposed DL-MAC protocol coexists with other nodes employing traditional protocols. A novel trust update mechanism for UANs based on RL is introduced in \cite{HJW2020}.

Recent studies of laser chaos have led to prior advantages in RL. Chaos in physics is closely associated with randomness and is extremely sensitive to initial conditions. In a diode laser emitter, chaos is generated by injecting delayed light into the cavity, which is achieved by redirecting the output of the semiconductor laser through a beam splitter towards a laser reflector. The resulting chaotic oscillation of the laser signal level can be used for various purposes such as secure communication, chaos-based sensing, and control.
Random numbers can be obtained by fast sampling of the signal level \cite{A2008}. Laser chaos-generated random numbers typically have a high frequency and have passed the National Institute of Standards and Technology random number test. Due to the superiority, research has been conducted on the potential applications of laser chaos. Naruse et al. \cite{NMH2018} proposed a multi-armed bandit algorithm using chaotically oscillating waveforms, which was further improved to consider the order correct probability of MAB \cite{NC2020}. Inspired by previous study applied for dynamic channel selection, this study is to solve the multi-sensor-nodes underwater acoustic relay assignment problem by using laser chaos \cite{THK2020}.

Although lots of work has been done for UANs with RL, most work only focuses on a single source node with highest throughput. To the best of the authors' knowledge, there are only a few existing studies that consider multiple source nodes. However, in the real underwater environment, it is crucial not only for each source node to choose the best relay but also to avoid collisions. Additionally, the exchange times are often limited in reality. Given the limitations of information exchange in reality, it is not possible to ensure a reward-maximizing solution in this system. Therefore, this paper no longer concentrates on the highest return but on stable allocation due to the particularity of multiple source nodes. Actually, this stable arrangement was originated from marriage matching models \cite{GS1962} and has been widely studied in many fields like channel selection, market resource allocation etc., and powerful performance has been proven in multichannel wireless networks \cite{LZY2012}. Furthermore, given that laser chaos has been proven to be effective in the aforementioned literature, it is worth investigating whether it is possible to achieve higher throughput by laser chaos to attain a suitable stable configuration in UANs.

From the above starting points, a novel stable relay assignment with laser chaos series for multiple source nodes, named laser chaos-based multi-processing learning (LC-ML) method, is proposed to enhance throughput and improve operational efficiency for UANs. This provides a practical and useful method for the relay assignment in a realistic UAN. To evaluate the effectiveness of the proposed method, several detailed comparisons are made with existing related results. First, previous work has focused primarily on achieving highest throughput with a single source node whereas this paper considers stable states with high throughput. Specifically, the ``high throughput" in our study is ensured through the definition of a stable arrangement (Definition 1 and 2), meaning it represents the optimal throughput that can be achieved while maintaining stability across the system. Second, unlike previous studies that have assumed fixed relay preferences, this study considers the ambiguous preferences of different relays in the world, and the relevant performance of the proposed method is evaluated by targeting this aim. Finally, this study applies random number generated by laser chaos in UANs, which is a unique and innovative approach to improving network efficiency.

\section{Underwater Acoustic Network}\label{sec2}

\subsection{UANs with Reinforcement Learning}

An underwater acoustic network (UAN) considered in this study consists of $K$ source nodes (SNs), $M$ mobile autonomous underwater vehicles (AUVs) and a base station (BS).  A conceptual diagram of this system has been shown in Fig. 1a. It can be seen that the BS, located at the sea surface, acts as the central hub for this network and receives information from SNs. SNs near the seabed typically serve as data collectors, gathering information about the underwater environment. Relay nodes, which are commonly located in the middle region of the UAN, serve as communication hubs. It facilitates the transfer of information between the SNs and the BS so that efficient communication can be achieved. Under this scenario, there has been widespread interest in finding the most appropriate acoustic modems for each SN in order to obtain higher data transmission. Without loss of generality, the assumption $M \leq K$ is taken in this study. Otherwise, the number of users transmitting information concurrently can be restricted through the use of split time to satisfy the condition $M \leq K$.

The working principle of the UAN can be understood approximately with a reinforcement learning problem (Fig. 1b). More detailed, each SN selects a relay and receives the corresponding feedback. In the subsequent time slot, the SN must decide whether to select a new relay with unknown performance or to maintain its current relay. This decision-making process parallels the classic exploration-exploitation trade-off in reinforcement learning. It is assumed that the SN is able to utilize only one relay per transmission slot. If multiple SNs utilize the same relay simultaneously, the information transmitted through the relay will be lost. Relays can communicate with each other to exchange relays.
The reward at a time slot $t$ is derived from an unknown Bernoulli distribution that is independent from the different relay. In other words, the return of the SN received from a relay is taken from a value of 0 or 1 with a certain probability. For any SN, denoted as $s$, and any AUV, denoted as $r$, the return of a successful transmission is denoted by $\mu_{s,r}$. Based on collision avoidance assumption, the total throughput of an UAN is defined as:
$$
R = \sum_{s = 1}^{K} \mu_{s,f(s)} \mathbb{I}(\nexists \;\text{SN} \;s'\neq s, s.t. \; f(s')=f(s)), $$
where $f(s)$ indicates the selected relay of SN $s$ and $\mathbb{I}$ is the indicator function.
Moreover, a general model is considered where probability vary for various SNs within the same relay. The probability that a relay exhibits a fully transmitted state is dependent on the SNs. This assumption is applicable in complex electromagnetic environments, where the environment changes according to distinct SNs.
\vspace{-0.3cm}
\begin{figure}[H]
  \centering
  \includegraphics[scale=0.5]{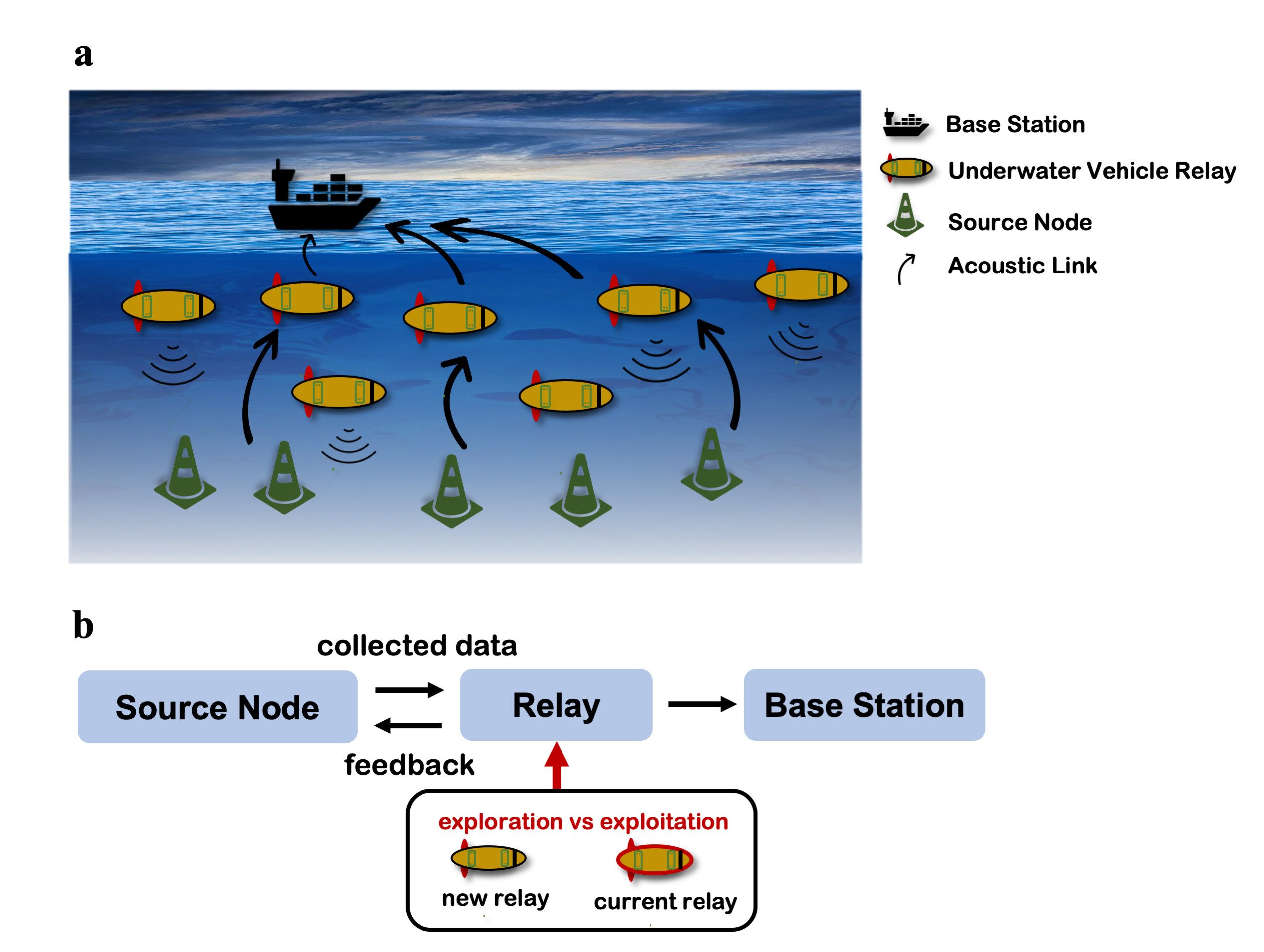}
  \caption{\textbf{An UAN scenario and data transmission process.} \textbf{a} An UAN with seven relays and five SNs. \textbf{b} The exploration and exploitation in an UAN.}
\end{figure}

\subsection{Stable Relay Matching Objectives}

Although existing work on relay selection in an UAN is mainly contributing on highest system data transmission, this study adopts two stable matching objectives, i.e., classical stable arrangement (CSA) and ambiguous stable arrangement (ASA) shown in Fig. 2.  On the one hand, in a realistic underwater environment, the attainment of an optimal maximum solution necessitates frequent exchange of information. However, the exchange times are often limited in reality and SNs are usually powered by batteries that cannot be recharged. Given these limitations, it is not possible to ensure a reward-maximizing solution in this system. On the other hand, this study focuses on multi-SNs with multi-AUVs in an UAN, which is not a simple analogy for a single SN. Collision challenges arise when multiple SNs simultaneously access the same relay and the primary purpose in this situation is to minimize the occurrences of collisions among SNs. Moreover, maximizing throughput in isolation could result in unstable configurations, where some SNs monopolize relay assignments, causing collisions or inefficient resource usage. Therefore, stability ensures that the network operates efficiently, with all participating nodes sharing the available resources fairly, avoiding congestion and conflicts.
 By accomplishing this goal, the network can operate more efficiently and effectively, leading to improved performance and better utilization of network resources. Therefore, the stable relay matching objectives are focused in the work.

It should be noted that the CSA objective has been investigated in many other fields such as marriage matching problem \cite{GS1962}, channel dynamic selection \cite{LZY2012}, but it has rarely been studied in an UAN system except for \cite{LLYH2018}. The objective of ASA is proposed to address the situation in which the cognitive differences between relays are uncertain. We assume that 

These relays follow adjacent Bernoulli distributions, and the ambiguity arises from the overlap in their characteristics. In these examples, even though switching to another relay may potentially result in a higher return, the relay may still choose to remain on the current one. This occurs when the average of the better relay is close enough to the current one and the cost of searching for a new relay and making a change may be excessively high. Based on these, CSA and ASA are defined as follows:
\begin{definition}
The Classical Stable Arrangement (CSA) is a mapping $f_C$ from the SN set to the AUV set that for any SN $s$ and AUV $r$ satisfying $\mu_{s,r}>\mu_{s,f_C(s)}(r\neq f_C(s))$, then there exists some other SN($s'$) such that $f_C(s')=r$ and $\mu_{s',r}>\mu_{s,r}$.
\end{definition}
\begin{definition}
The ambiguous stable arrangement (ASA) is a mapping $f_A$ from the SN set to the AUV set that for any SN $s$ and AUV $r$ satisfying the difference between $\mu_{s,r}(r \neq f_A(s))$ and $\mu_{s,f_A(s)}$ is less than a constant $c$, then there exists some SN $s'$ such that $f_A(s')=r$, the difference between $\mu_{s',f_A(s)}$ and $\mu_{s',r}$ is bigger than $c$ or the difference between $\mu_{s',r}$ and $\mu_{s,r}$ is bigger than $c$.
\end{definition}
\begin{figure}[H]
  \centering
  \includegraphics[scale=0.4]{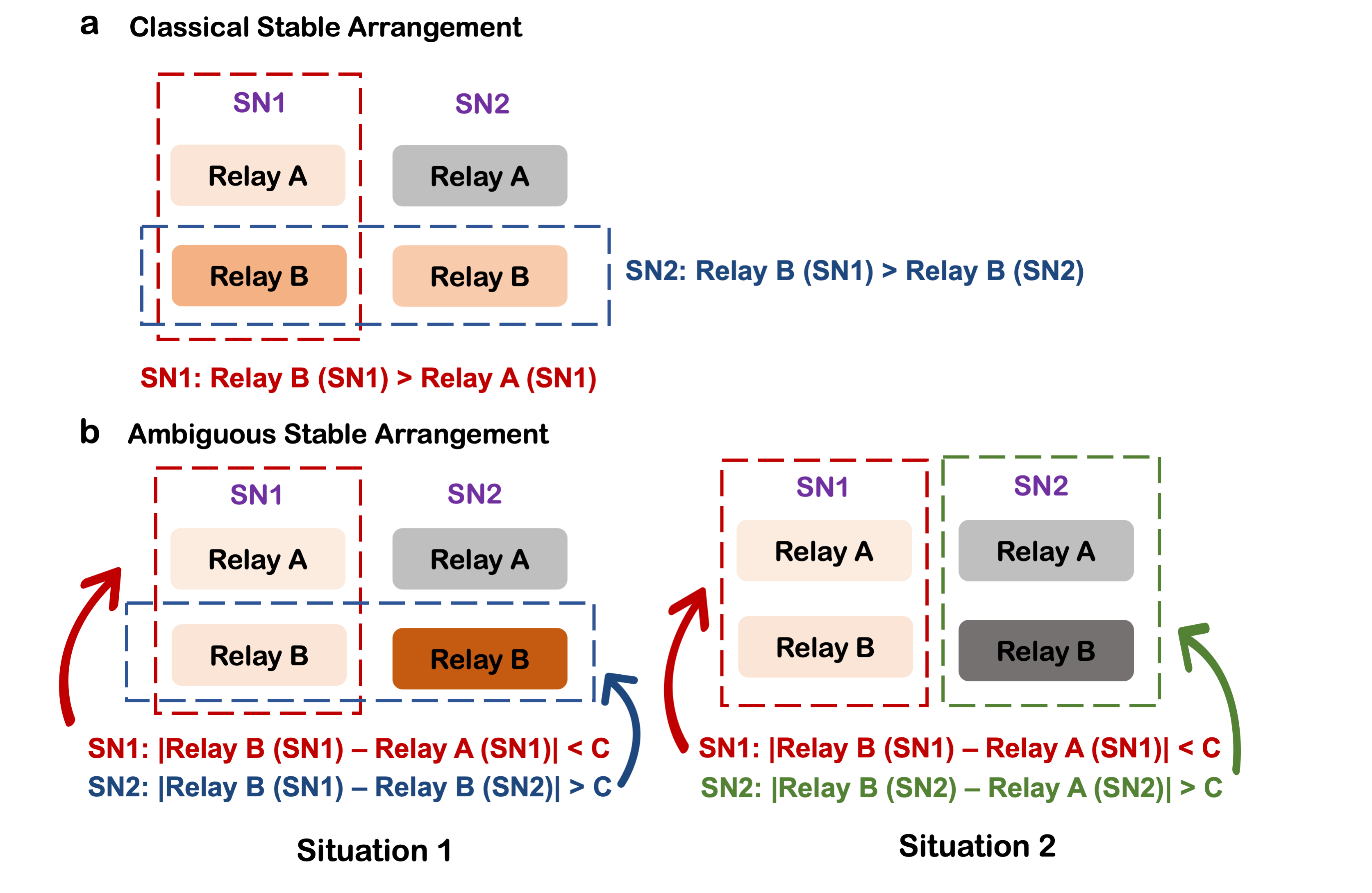}
  \caption{\textbf{The definitions of CSA and ASA.} \textbf{a} The conditions in classical stable arrangement where the deeper the color, the greater the corresponding return. \textbf{b} The situations in ambiguous stable arrangement where the deeper the color, the greater the corresponding return.}
\end{figure}

The key difference between CSA and ASA lies in their cognitive characterization. CSA is established with stability and clear preferences, while ASA is defined with uncertain or imprecise preferences. As shown in Fig. 2(a), in a CSA, if source node 1 (SN1) desires to exchange, there must be a source node 2 (SN2) that declines to do so, resulting in a stable state. More detailed, in this scenario, SN1 occupies relay A, while SN2 occupies relay B. To determine whether this state is stable according to CSA, we examine the situation where SN1 finds it preferable to switch to relay B because $\mu_{1,B}>\mu_{1,A}$. However, since $\mu_{2,B} > \mu_{1,B}$, SN2 is unwilling to exchange. Therefore, the system remains stable as no further exchange is possible under these conditions. Similarly, Fig. 2(b) illustrates two scenarios in an ASA where one of them can ensure efficient allocation.

\section{Laser Chaos-based learning}
\subsection{Laser Chaos Decision-making}

Random numbers can be generated rapidly using chaotic semiconductor lasers \cite{NMH2018}. The semiconductor laser is equipped with a polarization-maintaining (PM) coupler. The laser emits light that hits a variable fiber reflector, causing delayed optical feedback and inducing laser chaos. The output light at the end of the PM coupler is detected by a high-speed AC-coupled photo-detector after passing through an optical isolator (ISO) and an attenuator. The captured light is then sampled using a high-speed digital oscilloscope. It has been shown that the performance of random number generators can be greatly improved by using this kind of chaotic laser device \cite{J2012,A2012,NMH2018} and has been used for applications such as ultra-fast random number generation \cite{A2008,KARC2010} and decision-making \cite{Naruse2017,Naruse2018}.

Based on the these results, this study utilizes chaos-based techniques to update preferences which introduces a certain level of randomness and uncertainty into the system, allowing for the exploration of different options and the attainment of a more diverse range of solutions. The chaotic time series generated by the laser chaos method has unique features, such as adaptable randomness and negative autocorrelation, that make it superior to other alternatives for decision-making. This is supported by research \cite{Naruse2017} that has shown that chaotic dynamics in a laser system can result in superior decision-making abilities.

This study introduces a novel method called Laser Chaos-based Multi-processing Learning (LC-ML) for achieving two types of stable configurations, as illustrated in Fig 3. The findings demonstrate that LC-ML has the features of sufficient exploration.
Before proceeding with the exchange process, the system must first undergo an initial setup process, during which it determines the relays' preference sequence using laser chaos. At each time slot, the SN updates the preference list of each relay in the UAN system. The details of this process are described in Section III.C.

Subsequently, multiple requesters are randomly chosen during the exchange stage, and they will communicate with the relay that holds the best position according to the preference table. To achieve the two types of stable configurations, namely CSA and ASA, different conditions must be satisfied as definitions. These conditions ensure that the system reaches a state of stability. Additionally, the LC-ML method consists of two main steps: \textbf{Preference Updating} and \textbf{Multi-processing Exchange}, which will be elaborated upon in the following.
\begin{figure*}[!t]
\centering
   \setlength{\abovecaptionskip}{-0.2cm}
  \includegraphics[width=14cm, height=9cm]{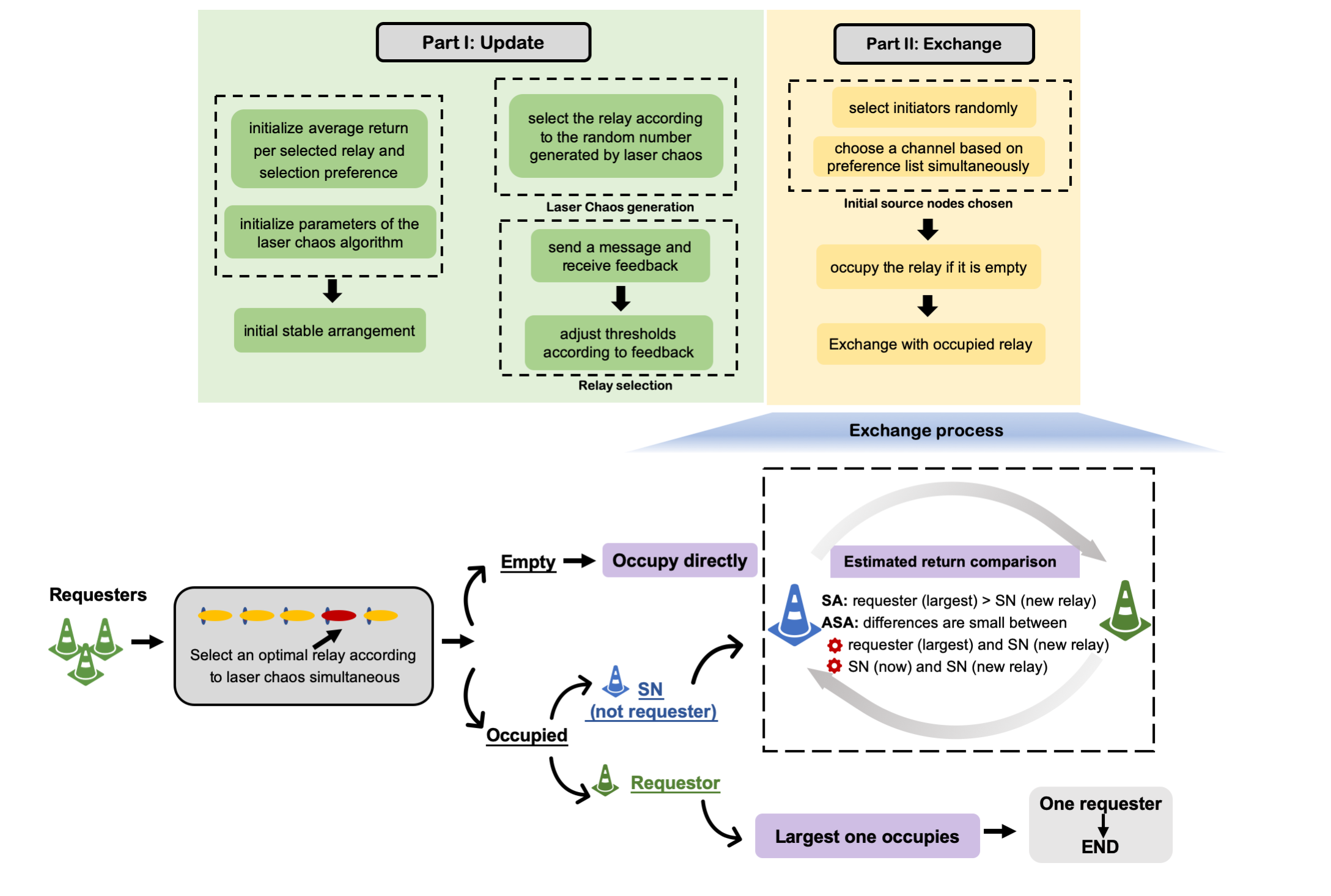}
  \caption{Overview of LC-ML and detailed exchange process.}
\end{figure*}

\subsection{Laser Chaos-based Preference Updating}
As a RL method, the process of learning potential options plays a pivotal role. This paper introduces the concept of utilizing laser chaos-generated random numbers, which differs from computer-generated pseudo random numbers. The methodology is adapted from \cite{Naruse2018,NB2016}. 

To illustrate the process of preference updating, we begin with the single SN case as an example. In the scenario of multiple SNs, each SN follows the same procedure. At each time slot $t$, relay selection is performed based on signals generated through laser chaos. Binary codes are used to uniquely identify all relays. Taking into account $M = 2^m$, each relay can be identified as $\{B_1 B_2\cdots B_m\}$, where $B_i \in \{0,1\}, i = 1,\cdots,m$. For example, if there are eight relays, they can be represented as $\{000,001, \dots, 110, 111\}$. If $M \neq 2^m$, there exists a positive integer $m'$ such that $2^{m'-1}<M <2^{m'}$. In this case, we set $m = m'$, and the binary codes $\{B_{1}B_{2} \ldots B_{m}\}$ are sequentially assigned to all $M$ channels. The remaining $ 2^{m}-M$ relays are treated as virtual relays, with a successful transmission probability of zero for all SNs. At every time slot $t$, the selection of a relay (represented by a binary code of length $m$) is determined through $m$ comparisons between the signal levels generated by laser chaos and predefined thresholds $C_{i,B_{1}...B_{i-1}}$, where $i= 1,\cdots,m$. The threshold $C_{i,B_{1}...B_{i-1}}$ denotes the value for determining $B_i$, given determined values of $B_1...B_{i-1}$. If the signal level exceeds the threshold $C_{i,B_{1}...B_{i-1}}$, then $B_i$ is set to 1; otherwise, it is assigned a value of 0.

Then SN $s$ sends a 1-bit information through the relay selected above. If the transmission by SN $s$ is successful, the threshold is updated as follows:
\begin{equation}
C_{i,B_{1}...B_{i-1}} = \alpha* C_{i,B_{1}...B_{i-1}} +\rho_1(-1)^{B_{i}},\;i= 1,\cdots,m. 
\end{equation}
Otherwise, it is updated as:
\begin{equation}
C_{i,B_{1}...B_{i-1}} = \alpha* C_{i,B_{1}...B_{i-1}} + \rho_{2}(-1)^{1-B_{i}},\;i=1,\cdots,m,
\end{equation}
where the definitions of $\rho_1$ and $\rho_{2}$ will be discussed below. 

While SN $s$ sends the 1-bit information, the selected relay’s features can also be learned. The variables $T_{t,s,r}$ and $S_{t,s,r}$ represent, respectively, the number of times relay $r$ was selected by SN $s$ before time $t$, and the number of successful transmissions choosing relay $r$ by SN $s$ before time $t$. Therefore, the selected time $T_{t,s,r}$ and the number of successful transmissions $S_{t,s,r}$ are updated, which determine the success rate of selecting relay $r$ by relay $s$ before time $t$:
\begin{equation}
\mu_{t,s,r}:=\frac{S_{t,s,r}}{T_{t,s,r}}.
\end{equation}
The value $\mu_{t,s,r}$ reflects the effectiveness of various relays for different SNs, serving as a crucial indicator for future relay selection decisions by SN $s$.

This step involves adjusting the threshold based on the feedback obtained in the preceding step. The feedback determines whether the threshold should be increased or decreased to raise or lower the probability of choosing relay $r$ in future iterations. The threshold adjustment is performed by adding or subtracting the parameters $\rho_1$ and $\rho_2$. In general, there are two types of the parameters $\rho_1, \rho_2$, namely fixed and flexible. For fixed parameters, both $\rho_1$ and $\rho_2$ remain constant at a value of 1. In contrast, for the flexible type, the values of $\rho_1$ is set to be 1 and $\rho_2$ is continuously modified according to \cite{KA2014} for $i=1,\ldots,m$:
\small{
\begin{equation} 
\rho_{2,i,B_{1}B_{2}...B_{i-1}} = \dfrac{\frac{T^{1}_{i,B_{1}B_{2}...B_{i-1},B_{i} =0}}{T^{2}_{i,B_{1}B_{2}...B_{i-1},B_{i} =0}}+\frac{T^{1}_{i,B_{1}B_{2}...B_{i-1},B_{i} =1}}{T^{2}_{i,B_{1}B_{2}...B_{i-1},B_{i} =1}}}{2-\Big(\frac{T^{1}_{i,B_{1}B_{2}...B_{i-1},B_{i} =0}}{T^{2}_{i,B_{1}B_{2}...B_{i-1},B_{i}=0}}+\frac{T^{1}_{i,B_{1}B_{2}...B_{i-1},B_{i}=1}}{T^{2}_{i,B_{1}B_{2}...B_{i-1},B_{i} =1}}\Big)},
\end{equation}
}where $T^{1}_{i,B_{1}B_{2}...B_{i-1},B_{i}=j}$ represents the number of successful transmission observed, and $T^{2}_{i,B_{1}B_{2}...B_{i-1},B_{i}=j}$ represents the number of times $B_i = j$ when $B_{1}B_{2}...B_{i-1}$ has been determined. To ensure optimal performance, it is crucial to select the appropriate type of parameters based on the characteristics of the relay distribution. This enables a more dynamic and adaptive threshold adjustment process, resulting in an efficient relay selection approach. This dynamic adjustment contributes to enhanced performance in various scenarios. The procedure is outlined as follows:

\begin{algorithm}[H]
\footnotesize{\caption{Preference Updating.}\label{alg:alg1}
\begin{algorithmic}
\STATE \textbf{for} each SN $s$ in SNs:
\STATE \hspace{0.5cm}\textbf{for} $i = 1,\cdots,m$:
\STATE \hspace{0.5cm}\hspace{0.5cm}\textbf{if}  signal level generated by laser chaos $>$ $ C_{i,B_{1}...B_{i-1}}$ \textbf{then}
\STATE	\hspace{0.5cm}\hspace{0.5cm}\hspace{0.5cm}$ B_{i} \gets 1$ 	
\STATE	\hspace{0.5cm}\hspace{0.5cm}\textbf{else}
\STATE \hspace{0.5cm}\hspace{0.5cm}\hspace{0.5cm}$ B_{i} \gets 0$
\STATE  \hspace{0.5cm}\hspace{0.5cm}\textbf{end if}
\STATE \hspace{0.5cm} \hspace{0.5cm}\textbf{return} selected relay and observe transmission result
\STATE  \hspace{0.5cm}\hspace{0.5cm} update the success rate, selection preference and parameters
\STATE	\hspace{0.5cm}\bf{end for}
\STATE	\bf{end for}
\end{algorithmic}}
\end{algorithm}

\subsection{Multi-processing Exchanging}
The primary objective of the exchange process is to establish a stable configuration where each SN is satisfied with its current relay choice and no collisions occur. Firstly, a set of SNs is randomly selected as requesters, who acts as the role of launching the request for exchange. Each requester simultaneously selects a relay based on its individual preference list. The selected relay then provides feedback to the requester about its current status, indicating whether it is unoccupied and only receiving the request from single requester. If the relay is unoccupied, the requester takes control of the relay immediately. However, if the relay is either occupied or is receiving requests from multiple requesters, further observations are necessary to proceed with the selection process. Considering two stable state in this paper, this article needs to provide two exchange processes.

\textbf{The process of CSA:}
When multiple requesters send requests to an unoccupied relay, the SN with the highest probability of using that relay to transmit occupies the relay and ends the loop, while other requesters continue the loop. When a relay is occupied by a specific SN, the probability of success between the occupying SN and the initiator under that relay is compared. The SN with the highest probability of success occupies the channel, and the others continue the loop.

\textbf{The process of ASA:}
The parallel switching process of ASA is mostly the same as that of CSA. However, when multiple requesters send requests to a relay $r_1$ occupied by SN $s_1$, if there exists a requester $r_2$ such that $|\mu_{r_2,r_1} - \mu_{r_1,s_1}| \leq c$ and $|\mu_{r_1,s_1} - \mu_{r_1,f(r_2)}| \leq c$, where $f(r_2)$ is the relay occupied by $r_2$, then $r_2$ occupies the $s_1$ and the original owner begins to loop. Otherwise, all requesters continue to the loop.

\begin{algorithm}[H]
\footnotesize{\caption{Multi-processing Exchange.}\label{alg:alg2}
\begin{algorithmic}
\STATE \textbf{Randomly select requesters:}
\STATE \hspace{0.5cm} \textbf{for} each requester:
\STATE	\hspace{0.5cm}\hspace{0.5cm}\textbf{if}
the selected relay is unoccupied or receives only one request:
\STATE	\hspace{0.5cm}\hspace{0.5cm}\hspace{0.5cm}occupy the channel
\STATE \hspace{0.5cm}\hspace{0.5cm}\textbf{else}
\STATE  \hspace{0.5cm}\hspace{0.5cm}\hspace{0.5cm} perform CSA or ASA exchange procedure
\STATE  \hspace{0.5cm}\hspace{0.5cm}\textbf{return} the relay to the requester
\STATE	 \hspace{0.5cm}\bf{end for}
\end{algorithmic}}
\end{algorithm}

\section{Throughput Performance}

It is important to pay attention on the throughput performance of this relay configuration learning method. While the primary goal of this method is to prevent collisions in the UAN system, achieving higher overall performance remains a critical objective. The overall performance will be measured by the ratio of successful transmission counts to the total number of trials, serving as a key factor in determining the effectiveness of the relay configuration learning method. The following discussions cover several aspects including comparisons of laser-chaos-generated random numbers to uniform distribution generated random numbers and Gaussian distribution generated random numbers, an evaluation of the overall performance of LC-ML in relation to prior work, the impact of varying numbers of requesters on performance, and the ability of the system to adapt to changing environments. In this work, we set $\alpha = 0.99$, $\rho_1 = \rho_2 = 1$, and all thresholds $ C_{i,B_{1}...B_{i-1}}, i =1,\cdots,m$ are set to 0.

\textbf{Laser-chaos-generated random numbers improve throughput.}
Naruse et al. \cite{NMH2018} have developed a method utilizing random numbers generated by the ultrahigh bandwidths of light waves and practical enabling technologies, which outperforms conventional methods, such as computer-generated pseudo-random numbers. The study adopted the preference updating methods similar to those described in \cite{NMH2018} and observed a positive impact on the overall performance of the relay configuration learning process. The results of testing different random number generation methods with varying numbers of requesters are shown in Fig. 4-5. The main objective of testing in Fig. 4 is CSA, while Fig. 5 focuses on ASA. The laser chaos method and the uniform distribution method significantly outperform the Gaussian distribution method. As a result, only the laser chaos and uniform distribution methods are compared in Fig. 5 with the goal of ASA, as the other methods perform poorly in CSA (Fig. 4). In some cases, the uniform distribution may initially perform better than the laser chaos method. This can be attributed to the fact that the random numbers generated by laser chaos are more unpredictable, resulting in more exploration when updating the preferences list. However, after a certain number of iterations, the laser chaos method is likely to outperform the others with a high probability.
\begin{figure*}[!t]
  \centering
   \setlength{\abovecaptionskip}{-0.2cm}
  \includegraphics[width=13cm, height=8cm]{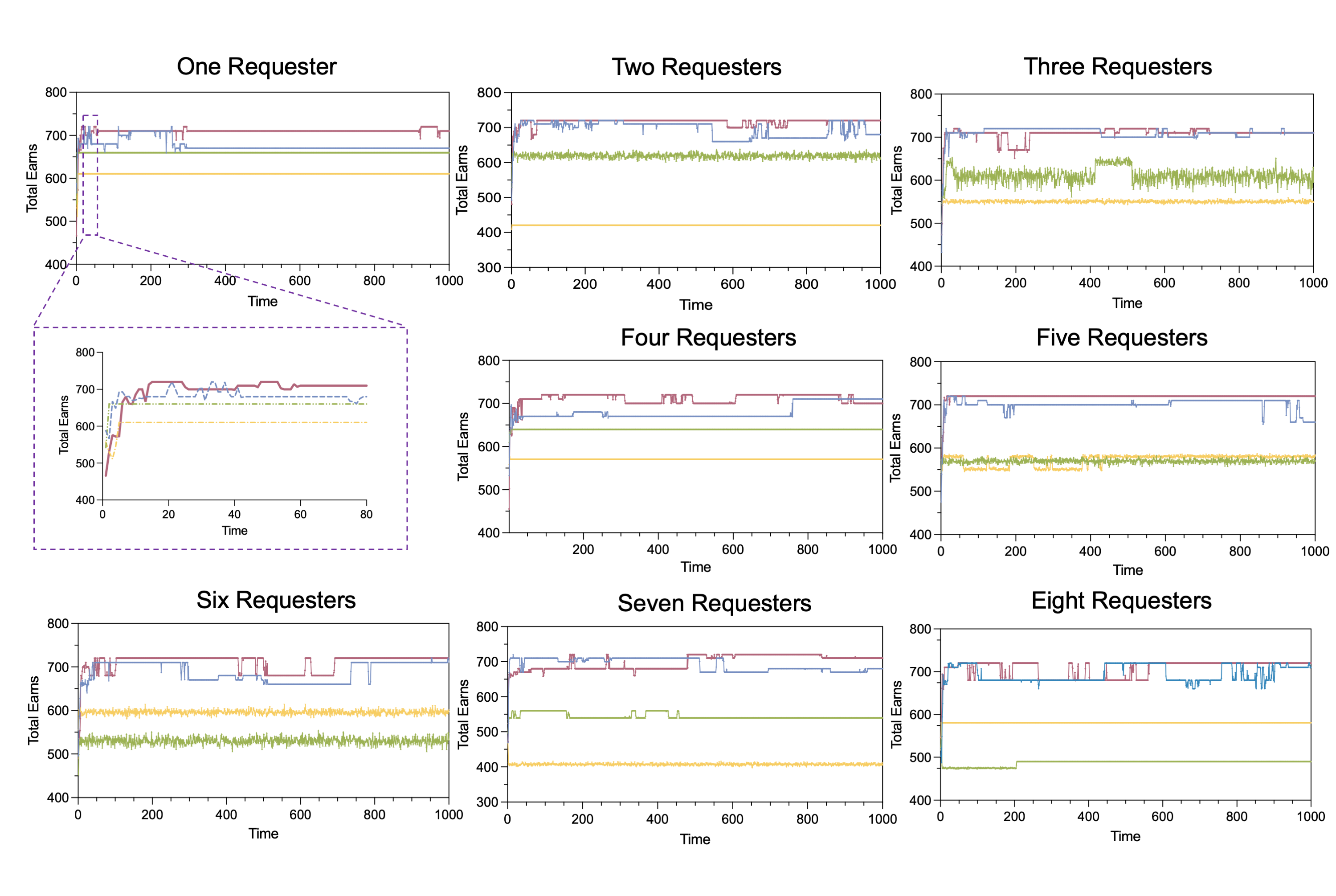}
  \caption{\textbf{Throughput comparisons between laser chaos-based and computer-generated random numbers for varying number of requesters with CSA.} The red line represents laser chaos-based random numbers,  the blue line represents uniform distribution random numbers, the green one is standard normal distribution random numbers and the yellow one is normal distribution random numbers with $a=1$ and $b=2$.}
\end{figure*}
\begin{figure*}[!t]
  \centering
   \setlength{\abovecaptionskip}{-0.2cm}
  \includegraphics[scale=0.45]{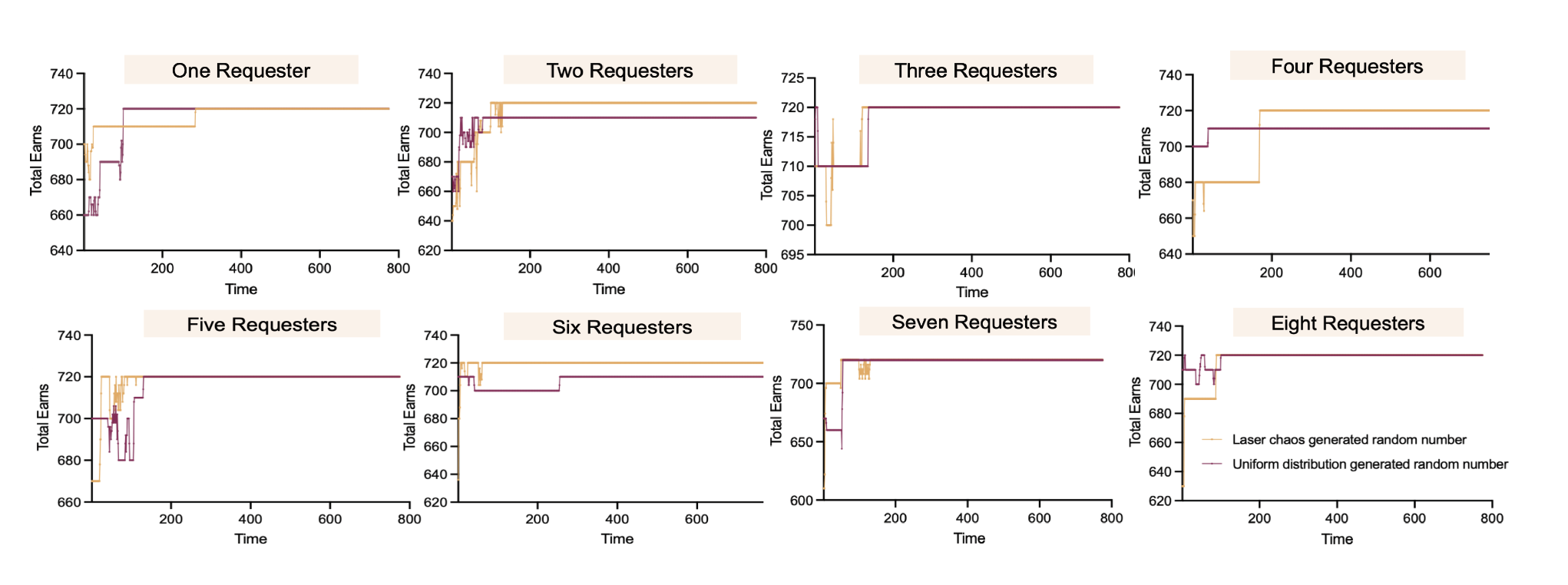}
  \caption{\textbf{Throughput comparisons with laser chaos based random numbers and normal distribution based random numbers for different number of requesters with ASA.}
}
\end{figure*}

\textbf{The LC-ML method also demonstrates good performance in terms of success transmission rate, with higher throughput achievable if ambiguous cognition is resolved.}
The stable configuration of a system is crucial, and a high throughput remains of significant importance. Despite the limited work conducted on stable configurations in UANs, this study compares the distributed stable matching multi-user (DSMU-MAB) approach introduced in \cite{LLYH2018} in Fig. 6, using five requester examples. The number of requesters are chosen on the basis of performance in Fig. 5. It can be found that the LC-ML method outperforms DSMU-MAB over the entire time slot.
Meanwhile, the performance of CSA and ASA is also tested in Fig. 6. This study finds that uncertain preferences may enhance communication, and the application of the LC-ML method in realistic UAN systems could have a positive impact compared to other learning methods. Moreover, the ASA curve has less fluctuation compared to CSA and this will also be observed in the subsequent experiment, indicating that ASA provides greater stability.
\begin{figure}[H]
  \centering
   \setlength{\abovecaptionskip}{-0.2cm}
  \includegraphics[scale=0.45]{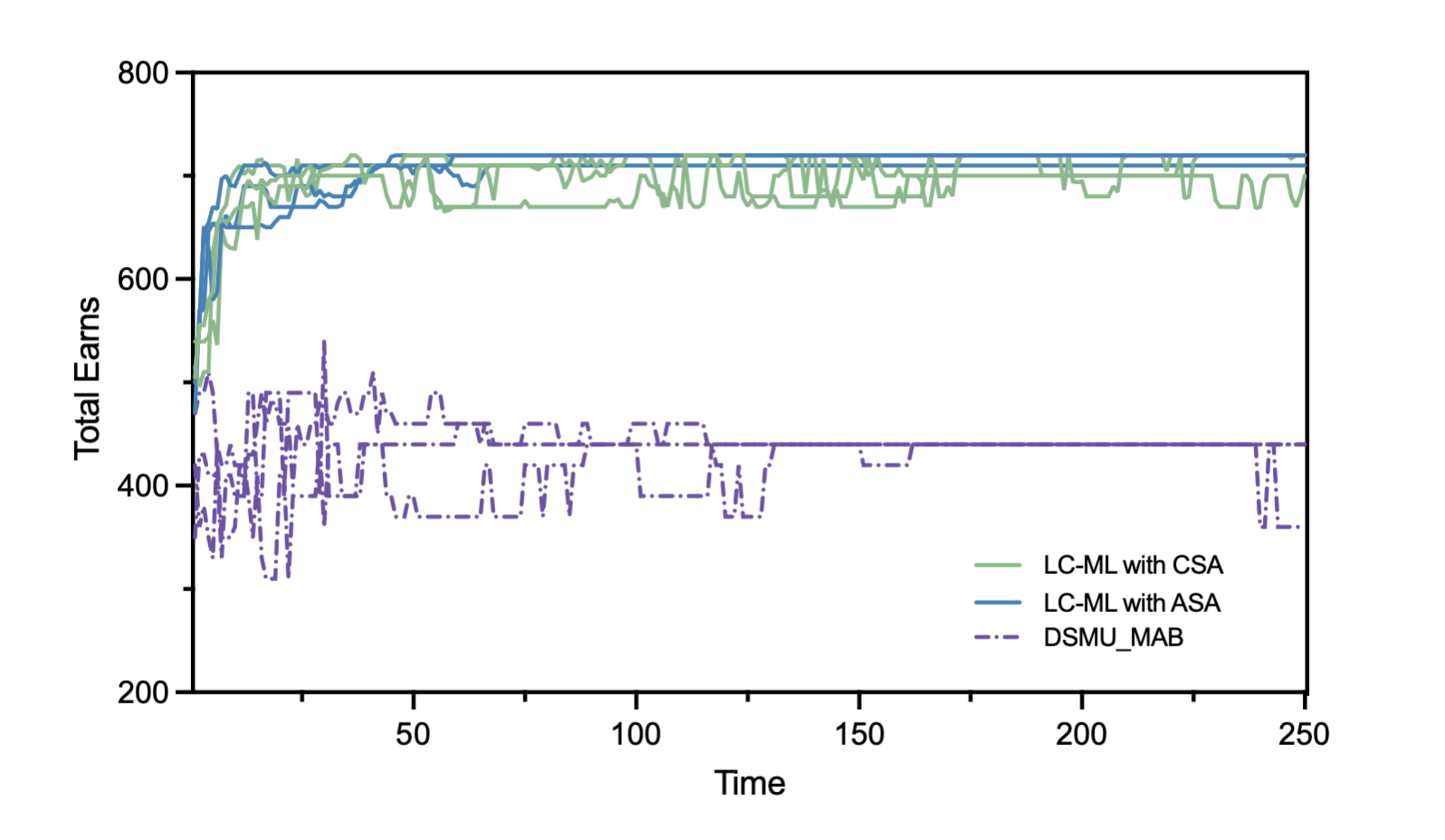}
  \caption{The performance of LC-ML with CSA, ASA and DSMU-MAB.}
\end{figure}

\textbf{The number of requesters has an influence on the throughput.} In multi-processing learning, multiple requesters are selected at the initiation of the communication process and launch an exchanging process on the basis of personal list. Exploring the effect of the number of requesters is crucial for maximizing the system performance. This can be achieved by conducting experiments with different numbers of requesters and evaluating the performance of the system in each configuration. By identifying the number of requesters that results in the highest performance, the system can be optimized for efficient handling of the workload. Moreover, testing with varying numbers of requesters can also provide insights into the scalability of the system and how it copes with increasing data and request volumes. Fig. 7 shows that different amounts indeed have an effect on the throughput and the color bar represents the colors corresponding to different numbers of initiators.

From the overall color changes, it is evident that multiple requesters tend to increase system output, as shown by the darker lines in the upper part of the figure.  It is still worth noting that this effect is much reduced in the case of ASA, which is due to the ambiguity of ASA. This ambiguity leads to a reduction in the number of exchanges, resulting in a lower volatility. After it reaches stable state quickly, the volatility starts to decrease compared to CSA and the curve approaches a plateau, as confirmed in the last experiment. However, it should be mentioned that due to the randomized nature of the random numbers used, determining the optimal number of requesters is challenging. Future work can focus on discovering the relationship between the optimal number of initiators and the number of users and repeaters.
\begin{figure}
  \centering
   \setlength{\abovecaptionskip}{-0.2cm}
  \includegraphics[scale=0.33]{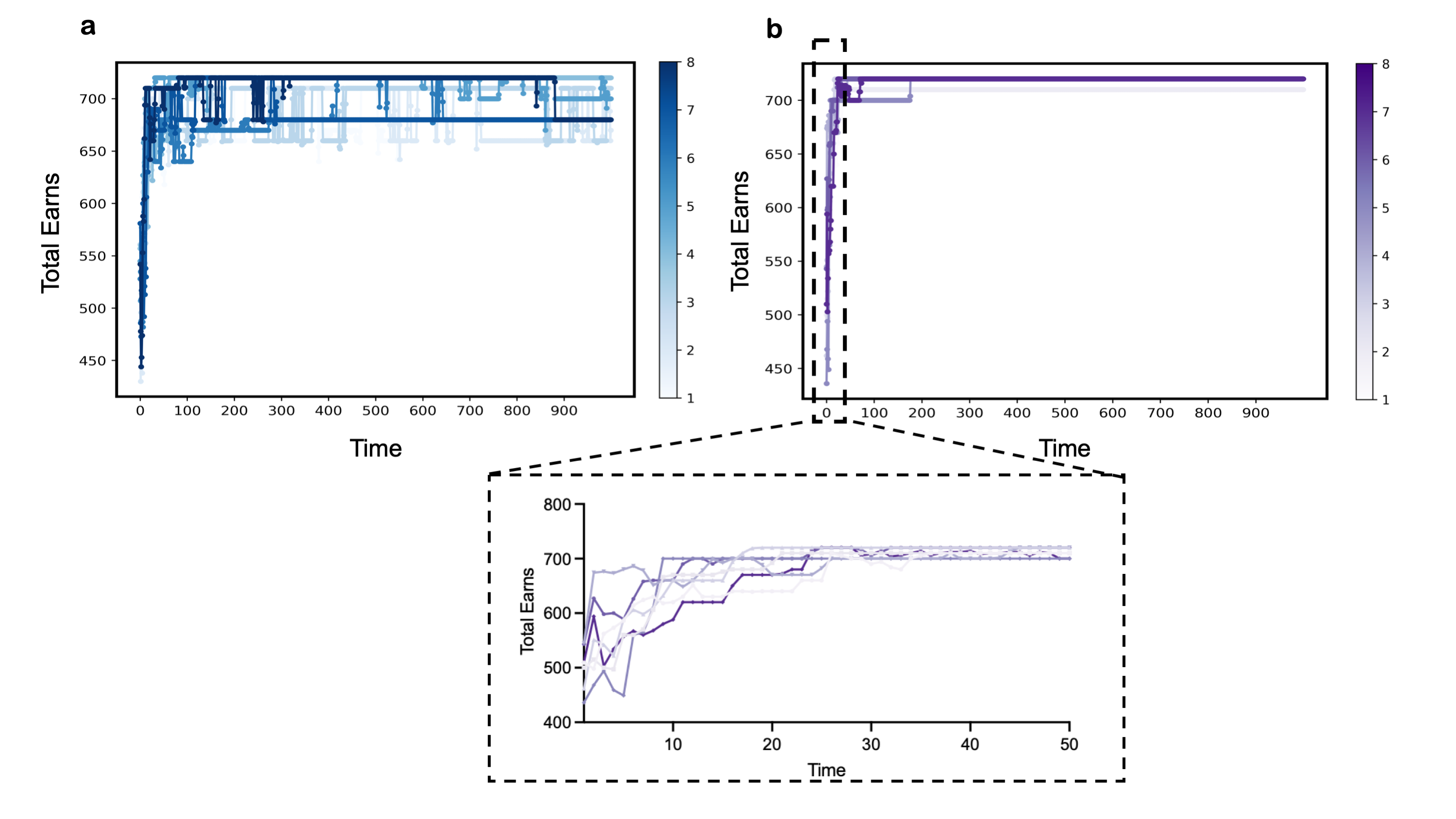}
  \caption{\textbf{The throughput of the system with different numbers of requestors.} \textbf{a} The situation with CSA. \textbf{b} The situation with ASA.}
\end{figure}


\textbf{The LC-ML method shows well in environmental adaptive ability.} In reality, the underwater environment, characterized by factors such as propagation delay and frequency-selective fading, is constantly changing. This necessitates the need for adaptable communication solutions and strategies to effectively operate in this dynamic environment. This poses a significant challenge in the design of communication systems for underwater environments, as they must be equipped to handle a wide range of conditions and variations. In light of this, this study assesses the capability of LC-ML when the environment changes (as shown in Fig. 8).

After 3,000 iterations, the relay's distribution was altered to simulate a new environment. As depicted in Fig. 8, the CSA was reached before the 3000th iteration. However, the information throughput experienced a significant decline after the 3000th iteration. To counteract this, the channel exchange process was restarted, allowing for the re-establishment of the CSA under the new distribution. Moreover, the number of channel exchanges, which had previously remained constant, also increased following the 3000th iteration.
\begin{figure*}
  \centering
   \setlength{\abovecaptionskip}{-0.2cm}
  \includegraphics[width=17cm, height=8cm]{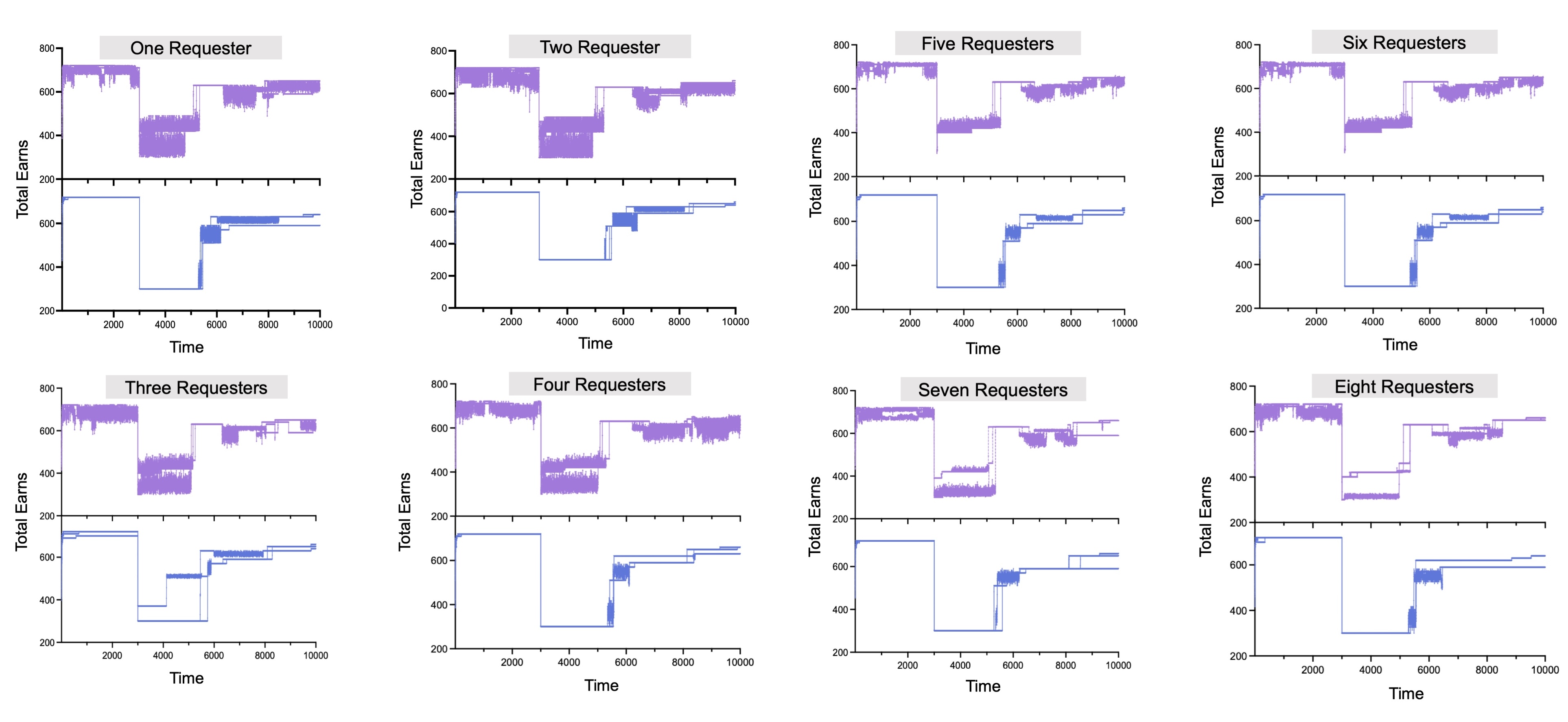}
  \caption{The UAN throughput with changed environment for multi-requesters.}
\end{figure*}

\section{Discussion}

In this study, the problem of acoustic relay assignment using laser chaos in an UAN has been investigated and this may be the first time to consider stable configuration in UANs with laser chaos. In order to explain the meaning of stable matching, two relevant definitions (CSA and ASA) are proposed, which are inspired from classical marriage stable matching and inaccurate cognitive preferences. Based on those aims, this UAN system reaches an equilibrium with multi-SNs cooperating simultaneously rather than pursuing individual highest throughput as commonly found in the  existing literature. Furthermore, a LC-ML method is designed to efficiently address these two objectives. This method is equipped with the randomness properties of laser chaos and multi-exchanging processes, resulting in higher throughput and strong adaptability with environmental changes over time.

It should be noted that the UAN considered in this study is not dynamic and complicated. Further research will be required to investigate the UAN scenario closer to practical underwater environment and evaluate potential claims of advantage. Moreover, comparing not only throughput but also network sustainability, such as the lifetime of AUVs, would provide a more comprehensive evaluation. In conclusion, despite some limitations, we believe that the tools developed in this work offer valuable insights and pave the way for future advancements in this exciting field.

\section*{Acknowledgment}

The authors would like to thank the editor and two anonymous referees for providing extensive and constructive suggestions that significantly improved this paper. The work is supported by the National Key R\&D Program of China (grant No. 2018YFA0703900), the National Natural Science Foundation of China (Grant Num. 11901352) and the Shandong Provincial Natural Science Foundation (No.ZR2019ZD41, No.ZR2021MA098).

 \bibliography{sn-bibliography}

\vfill

\end{document}